\documentclass[conference]{IEEEtran}
\IEEEoverridecommandlockouts

\usepackage[utf8]{inputenc}
\usepackage[english]{babel}

\usepackage{booktabs}
\usepackage{rotating,tabularx}
\usepackage{multirow}
\usepackage{multicol}
\usepackage[table,pdftex,dvipsnames]{xcolor}
\usepackage{color, colortbl}
\usepackage[first=0,last=9]{lcg}
\definecolor{LightGray}{rgb}{0.7,0.7,0.7}

\usepackage[caption=false,font=footnotesize]{subfig}
\usepackage{graphicx}
\graphicspath{{./figure/}}

\usepackage{bm}
\usepackage{amsfonts}
\usepackage{amsmath}
\usepackage{amssymb}
\usepackage{amsbsy}
\usepackage{mathrsfs} 
\usepackage{mathabx}
\DeclareMathSymbol{\shortminus}{\mathbin}{AMSa}{"39}

\usepackage{verbatim}
\usepackage{blindtext}
\usepackage{tcolorbox}
\usepackage{scalerel}
\usepackage{url}
\usepackage{adjustbox}
\usepackage{soul}
\usepackage{cite}
\usepackage{tikz}
\usepackage{algpseudocode}

\usepackage[wby]{callouts}
\usetikzlibrary{shapes.multipart}

\usepackage{array}
\usepackage{makecell}

\allowdisplaybreaks
\usepackage{amsthm}

\usepackage[linesnumbered, ruled]{algorithm2e}

\SetKw{Continue}{continue}
\SetKw{And}{and}
\SetKw{Elif}{elif}
\SetKw{True}{True}
\SetKw{False}{False}
\SetKw{To}{to}
\SetKw{Def}{def}
\SetKw{Not}{not}

\SetKwComment{Comment}{\color{green!50!black}// }{}

\newcommand{\FuncCall}[2]{\texttt{\bfseries #1(#2)}}
\SetKwProg{Function}{function}{}{}

\newcommand{\algosize}{\small}  % network size
\makeatletter
\renewcommand{\@algocf@capt@plain}{above}% formerly {bottom}
\makeatother
\SetAlCapNameFnt{\small}
\SetAlCapFnt{\small}
\newcommand{\R}{\mathbb{R}}
\newcommand{\Z}{\mathbb{Z}}

\newcommand{\G}{\mathcal{G}}
\newcommand{\V}{\mathcal{V}}
\newcommand{\E}{\mathcal{E}}

\usepackage{pifont}% http://ctan.org/pkg/pifont
\newcommand{\cmark}{\ding{51}}%
\newcommand{\xmark}{\ding{55}}%
\newcommand{\NA}{--}

\makeatletter
\patchcmd{\@maketitle}
{\addvspace{0.5\baselineskip}\egroup}
{\addvspace{-1\baselineskip}\egroup}
{}
{}
\makeatother

\begin{document}

	\title{Generating Connected, Simple, and Realistic Cyber Graphs for Smart Grids\\
	\thanks{This work was supported by NSF under Award Number 1808064.}
	}
	
	% \author{Osman Boyaci, Mohammad Rasoul Narimani, Katherine Davis, and Erchin Serpedin}
	\author{
		\IEEEauthorblockN{Osman Boyaci}
		\IEEEauthorblockA{
			Electrical Engineering\\
			Texas A\&M University\\
			College Station, TX, 77843\\
			osman.boyaci@tamu.edu
		}  \and
		\IEEEauthorblockN{M. Rasoul Narimani}
		\IEEEauthorblockA{
			College of Engineering\\
			Arkansas State University\\
			Jonesboro, AR, 72404\\
			mnarimani@astate.edu
		} \and
		\IEEEauthorblockN{Katherine Davis}
		\IEEEauthorblockA{
			Electrical Engineering\\
			Texas A\&M University\\
			College Station, TX, 77843\\
			katedavis@tamu.edu
		} \and
		\IEEEauthorblockN{Erchin Serpedin}
		\IEEEauthorblockA{
			Electrical Engineering\\
			Texas A\&M University\\
			College Station, TX, 77843\\
			eserpedin@tamu.edu
		}
		% \\[-10ex]
	}
	
	\maketitle

	\begin{abstract}
	Smart grids integrate communication systems with power networks to  enable  power grids  operation and command through real-time data collection and control signals. 
	Designing, analyzing, and simulating smart grid infrastructures as well as predicting the impact of power network failures strongly depend on the topologies of the underlying power network  and communication system. 
	Despite the substantial impact that the communication systems bring to smart grid operation, the topology of communication systems employed in smart grids   was less studied. 
	The power community lacks realistic generative communication system models that can be calibrated to match real-world data.
	To address this issue, this paper proposes a framework to generate the underlying topological graphs for the communication systems deployed in smart grids by mimicking the topology of  real-world smart grids.  
	In this regard, we have updated the Chung-Lu algorithm to guarantee  the communication network connectivity and to match the degree distribution of a real-world smart grid rather than following an expected degree distribution. In addition, 
	key characteristics of communication systems such as  diameter, average shortest paths, clustering coefficients, assortativity, and spectral gap were taken into consideration to generate the most similar real-world communication network for smart grid studies. We believe that the proposed algorithm to generate realistic cyber graphs for smart grid studies will benefit the power community.
	\end{abstract}

	\section*{Nomenclature}
	\addcontentsline{toc}{section}{Nomenclature}
	\begin{IEEEdescription}[\IEEEusemathlabelsep\IEEEsetlabelwidth{$12345678$}]
		\item[$\G,\V, \E$] Graph, set of nodes, set of edges
		\item[$n, \ m \in \Z$] $|\V|, |\E|$
%		\item[$u \in \V$] A node
%		\item[$(u,v) \in \E$] An edge
		\item[$u.d \in \R$] Degree of a node $u$
		% \item[$U \in \R^n $] 1D Vector
		\item[$S \in \Z^n $] Degree sequence
		\item[$\overline{d} \in \Z$] Max. degree
		% \item[$\underline{d}$] Min. degree
		% \item[$k_d \in \Z$] Number of nodes with degree $d$
		\item[$\widehat{(\cdot)}$] Normalization operation s.t. $ \sum_{1}^{n} \widehat{(\cdot)} = 1$
		% \item[$\widehat{U} \in \R^n $] Normalized vector s.t. 
		\item[$D \in \Z^{\overline{d}}$] Degree vector $D = [1, \dots, \overline{d}]$
		% \item[$K \in \Z^{\overline{d}}$] Count vector $K = [k_{1}, \dots, k_{\overline{d}}]$
		% \item[$F \in \R^{\overline{d}} $] A frequency vector $F = [\frac{k_{1}}{2m}, \dots, \frac{k_{\overline{d}}}{2m}]$
		\item[$ \widehat{K} \in \R^{\overline{d}} $] Normalized frequency vector %$F = \widehat{K}$
		\item[$\G.g \in \{0,1\}$] $\G$ is graphical (no self loops and parallel edges)
		\item[$\G.c \in \{0,1\}$] $\G$ is connected
		\item[$\rho \in \R$] Density as ${m} / {n}$
		\item[$\diameter \in \R$] Diameter as the longest shortest path
		\item[$\widetilde{sp} \in \R$] Average shortest path
		\item[$cc \in \R$] Clustering coefficient
		\item[$a \in \R$] Assortativity as the Pearson correlation coefficient of degrees between pairs of linked nodes
		\item[$\lambda \in \R$] Spectral gap as the minimum non-zero eigenvalue of the normalized Laplacian
		% \item[$f_{\theta}(x)$] Function with parameter $\theta$ and input $x$
	\end{IEEEdescription}

	\section{Introduction} \label{sec:intro}
	
    Communication systems play a major role in the deployment of smart grids empowering them to be more resistant, secure, reliable and manageable and ensuring connectivity of the grid components. 
    The backbone of communication systems in smart grids is represented by the information and communication technologies that allow two-way communication and automated control. 
    Communication systems improve the efficiency and reliability of smart grids by gathering and transmitting a wide variety of data  for grid control and decision-making purposes.
    The integration of cyber communications and control systems into the power distribution infrastructure has a profound impact on the operation, reliability, and efficiency of the power grid.
    The power and communication systems in modern power grids are highly intertwined.  Analyzing, simulating, designing, and predicting the impact of network failures strongly rely on the knowledge of a communication network topology~\cite{scaglione}.
    Thus, studying the underlying communication network topology is essential for the smart grid operation and control~\cite{PNNL_Report}. 
    
    In spite of the many  models proposed for electrical power systems~\cite{pglib}, the problem of modeling the underlying communication network in smart grids was less studied. In fact,  despite the huge efforts deployed for studying smart grids operation and control, modeling  smart grids is still at its infancy. There is not enough realistic and practical information about the topology of the underlying communication network in smart grids.
    So far, various efforts have focused on developing cyber-physical test models for general use by the power system community~\cite{kate1, kate2, kate3, kate4}.
    These studies consider different characteristics of communication systems including vulnerabilities of communication devices, attack paths, etc., to design a practical cyber layer for cyber-physical power systems. However, 
    taking all these characteristics into consideration makes these approaches computationally intractable for larger cyber graphs as the number of attack paths increases exponentially with the number of nodes. 
    
    To analyze the impact of cyber-graphs on power networks operation, e.g., cascading failure analysis, we  first need a fast and reliable framework to generate realistic cyber graphs for power test cases irrespective of their size. 
    A  few efforts were conducted for generating realistic cyber graphs for power test cases. 
    A graph generator based on the characteristics of Luxembourg smart grid, which is a power-line communication (PLC) system~\cite{Galli}, was presented in~\cite{scaglione} to create random but realistic smart grid communication topologies.
    Different characteristics of power grid including nodal degree distribution, graph spectrum and connectivity scaling property, etc., were analyzed in~\cite{Wang} for designing efficient communication schemes for power test cases.
    Heuristic algorithms were employed in~\cite{Kounev} to improve the communication reliability for smart grids at the transmission level. 
    However, many generic graph generation algorithms such as configuration model \cite{newman2003structure}, Havel-Hakimi algorithm \cite{hakimi1962realizability}, and Chung-Lu algorithm do not guarantee the graphical and connected graphs properties, and thus are not fit for designing communication systems for smart grids.
    Horv\'{a}t-Modes model, on the contrary, yields both connected and graphical outputs. 
    Due to its edge connection mechanism it produces graphs with large diameter and low assortativity, which are not realistic for communication systems. 
    Thus, it is important to propose  generative graph algorithms that take into account the real-world characteristics of communication systems in the design process.   
    
    Power system statistics were leveraged to design optimal communication systems for smart grids in \cite{Wang}.
    Similarly, the communication system statistics can be leveraged for designing a realistic communication system for smart grids  which is the centerpiece of this paper.
    Along this parallel, we first derive the statistical metrics of a real-world smart grid's communication system and then propose a graph generator based on the statistical information of the smart grid's communication system.
    Different graph attributes of a real-world smart grid communication graph such as diameter, assortativity, etc., are taken into consideration in designing the communication system for power test cases.
    Moreover, we adapt the Chung-Lu algorithm to preserve the connectivity of the graph since the connectivity is a key  characteristic of communication systems.
    In addition, edge switching operation is employed to prevent self loops and parallel paths in the communication graph.
    
	The contributions of this work are outlined as follows:
	% \begin{itemize}
		% generated graph is simple connected graph.
		(1) To generate connected, simple, and realistic cyber graphs for smart grids, we propose a simple and elegant framework by updating the Chung-Lu algorithm. 
		% generated graph degree distribution exactly matches the required degree distribution
		(2) To satisfy the required degree distribution, we propose an adaptive remaining degree approach instead of the fixed expected degrees.
		% generated cyber graph minimizes the cross edges between power and cyber layer
		(3) To minimize the length of cross edges between power and cyber graphs, we employ the Hungarian algorithm for optimal matching between cyber and power nodes.	
		% comparison
		(4) To compare the proposed method with the currently available approaches, we implement other graph generation methods in the literature such as configuration model, Havel-Hakimi, Horv\'{a}t-Modes, and Chung-Lu algorithms using the same degree sequence and analyze the global characteristics of the output graphs. 	
	% \end{itemize}

	The remainder of this paper is divided into four sections.
	Section II proposes the cyber graph generation framework.
	Section III presents the generated graphs and discusses their global characteristics.
	Finally, Section IV concludes the paper. 

	\section{Cyber-Graph Generation} \label{sec:gen}
	\subsection{Analyzing a real-world communication system}
	To be able to generate realistic cyber graphs, we first analyze a real communication system of a smart grid given in \text{\cite[Table~(3.2)]{PNNL_Report}}.
	Its degree count vector $K^*$ is extracted as $K^* = [162, 101, 30, 25, 11, 4, 3, 4, 2, 1]$ and its global graph characteristics are tabulated in Table~\ref{tab:ref_char}.
 	\begin{table}[h!]
		\centering
		\newcolumntype{?}[1]{!{\vrule width #1}}
		\setlength{\tabcolsep}{5.0pt}
		\renewcommand{\arraystretch}{1.4}
		\caption{Global graph characteristics of the reference system \text{\cite[Table~(3.1)]{PNNL_Report}}}
		\begin{tabular}{c | c | c | c | c | c | c | c}
			$|V|$ & $|E|$ & $\rho$ & $\diameter $ & $\widetilde{sp}$ & $cc$   &  $a$  & $\lambda$ 	     \\ \specialrule{1pt}{1pt}{1pt} 
			343  & 357   & 1.04   & 28           & 11.47     & 0.05  & -0.22 & $1.16e^{-3}$ \\    
		\end{tabular}
		\label{tab:ref_char}
	\end{table}
	It is clear from the distribution of $K^*$ that the reference graph shows scale free topology since $K^*$ values tend to diminish for larger degrees ($K^*_{1} = 362$  \(\gg\) $K^*_{10} = 1$).
	In addition, its global characteristics indicate that it is a sparse ($\rho=1.04$) and tree like graph (relatively high $\diameter$ and $\widetilde{sp}$).
	Moreover, it has only a few cyclic structures ($cc=0.05$), tends to connect higher degree nodes with a lower degree ones ($a=-0.22$) and presents  some bottleneck edges ($\lambda=1.16e^{-3}$) that removal of them may split the graph into different components\cite{PNNL_Report}. 
	
	\subsection{Selecting the degree distribution function}
\begin{figure}[h!]
		\centering
		\includegraphics[width=0.40\textwidth]{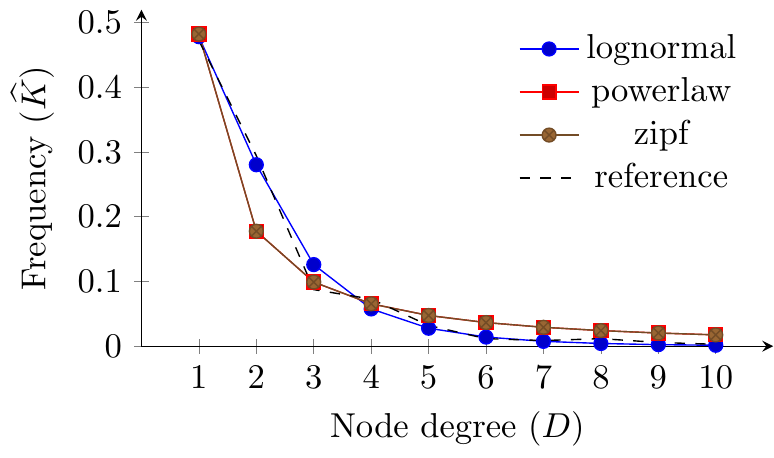}
		\caption{Normalized frequencies of the optimized degree distribution functions to approximate the reference distribution}
		\label{fig:dist}
	\end{figure}

	To select an appropriate degree distribution function that can generate the required degree distribution, we consider and optimize the parameters of three main distribution: generalized lognormal distribution \cite{kolda2014scalable}, power law distribution \cite{cristelli2012there} and zipf distribution \cite{cristelli2012there}.
	Specifically, we first obtain the frequency vector $\widehat{K^*}$ of the reference system and optimize the parameters $\theta$ of each distribution function $f$ by:
	\begin{equation} \label{eq:opt_dist}
		\min_{\theta} || \widehat{K^f} - \widehat{K^*} ||_2  , 
	\end{equation}
	where $\widehat{K^*}$ is the normalized frequency vector of the reference system and $\widehat{K_f}$ denotes the frequency vector of $K^f$ with $K^f_d = f_{\theta}(d)$.
	% for $ \forall d \in D$
	% $K = [f_{\theta}(1), \dots, f_{\theta}(\overline{d})]$    $F_k = $
	As can be seen from  Fig.~\ref{fig:dist} while powerlaw and zipf distributions underestimate  $\widehat{K^*}$ when $D<4$ and overestimates it when $D>4$, 
	lognormal distribution better approximates the reference distribution for all $D$ values.
	Formulation of the distributions, their optimal parameters, and mean square errors when estimating  $\widehat{K^*}$ is given in Table~\ref{tab:dist}.
	Note that since the lognormal distribution provides the best approximation, we use the lognormal distribution $f^*(x) = f_{\alpha=1.371, \beta=1.986}(x)$ to generate the degree sequences for the rest of the paper.
	\begin{table}[h!]
		\centering
		\newcolumntype{?}[1]{!{\vrule width #1}}
		\setlength{\tabcolsep}{4pt}
		\caption{Optimized degree distribution functions}
		\renewcommand{\arraystretch}{1.4}
		\begin{tabular}{ c ?{1pt} l | c | c | c }
			\textbf{distribution} & \textbf{formula} & $\bm{\alpha}$ & $\bm{\beta}$  & \textbf{MSE} \\ \specialrule{1pt}{1pt}{1pt}
			lognormal    & $f_{\alpha,\beta}(x) = \textrm{exp}\Big(\shortminus \big(\frac{log(x)}{\alpha}\big)^\beta\Big)$ & 1.371 & 1.986 & $2e^{-4}$  \\ \hline
			powerlaw     & $g_{\alpha,\beta}(x) = \beta x^{-\alpha}$ 							& 1.440 & 3.745 & $1.6e^{-3}$\\ \hline
			zipf         & $h_{\alpha}(x) = x^{-\alpha} / \zeta(\alpha)$ 					& 1.440 &       & $1.6e^{-3}$\\
		\end{tabular}
		\label{tab:dist}
	\end{table}

	\subsection{Generating valid degree sequence}
	The first step for generating a valid degree sequence for a realistic cyber graph is to specify the  number of nodes, $n$, number of edges, $m$, and maximum degree for each node, $\overline{d}$, \cite{PNNL_Report}.
	Then, we can generate the degree sequence $S$ by randomly drawing $n$ samples from the degree vector $D = [1, \dots, \overline{d}]$ with the corresponding probabilities $P(d=\widehat{F}_{d})$ where $F = [f^*(1), \dots, f^*(\overline{d})]$.
	Note that not every $S$ is realizable since it should satisfy $\sum_{i=1}^{n}S_i = 2m$ for an undirected graph.
	In addition, to be able to generate a simple graph without any self-loops or parallel edges, $S$ should be graphical.
	For instance, $S1=[1,1,4]$ and $S2=[2,2]$ are not graphical due to a self-loop and a parallel edge, respectively.
	To test whether a given degree sequence $S$ is graphical or not, we utilize the well-known Havel Hakimi algorithm \cite{hakimi1962realizability} given in Algorithm~\ref{alg:check}. If $S$ is not graphical, a new degree sequence should be generated.
	{\SetAlgoNoLine
	\begin{algorithm}[h!] % check
		\SetAlgoNoLine
		\algosize
		\caption{Check if the degree sequence is graphical}
		\label{alg:check}
		% config
		\SetKwInOut{Input}{Input}
		\SetKwInOut{Output}{Output}
		\SetKwFunction{FMain}{Main}
		\SetKwFunction{FGen}{Generate}
		\DontPrintSemicolon		
		\Function{is\_simple\_graph($S$)}{
			\While{\True}{
				sort S ascending \;
				$n \gets |S|$ \;  
				% Check if all the elements are equal to 0
				\lIf{$S_1 = 0$ $\And$ $S_n = 0$}{
					\Return \True}
				% Store the first element in a variable and delete it from the list
				$v \gets \ $\FuncCall{popleft}{$S$} \;
				% Check if enough elements are present in the list
				\lIf{$v > n-1$}{
					\Return \False
				}
				% Subtract first element from next v elements
				\For{$i \gets 1 \ \To \ v$}{%
					decrement $S_i$
					\lIf{$S_i < 0$}{
						\Return \False
					}
				}
			}
		}
	\end{algorithm}
	}

	\subsection{Generating simple connected cyber graph}
	The main algorithm outlined in Algorithm~\ref{alg:main}, for generating a simple graph $\G$ from a degree sequence $S$, presents five basic steps.
	{\SetAlgoNoLine
		\begin{algorithm}[h!] % main
			\algosize
			\caption{Generate a simple connected graph (main)}
			\label{alg:main}
			\DontPrintSemicolon
			\Function{main($S$)}{
				$n, m \gets |S|, sum($S$)/2$  \;
				create nodes $ {v_1,v_2,\ldots, v_n}$ s.t. $v_i.d \gets S[i]$ \;
				$I, J \gets \{ \ \}, \{v_1,v_2,\ldots, v_n\}$  \;	
				create $\G$ s.t. $\G.\V = \{ \ \}, \G.\E =  [ \ ]   $					\;
				$v_{max} \gets $ max. degree node \;
				add $v_{max}$ to $\G.\V$ as the first node \;
				remove $v_{max}$ from $J$ and add it to $I$ \;
				% init is done
				\FuncCall{generate\_tree}{$\G, I, J, n$} \;
				$E_s, E_p \gets$ \FuncCall{add\_remaining\_edges}{$\G, I, n, m$} \;
				\FuncCall{remove\_self\_loops}{$\G, E_s$} \;
				\FuncCall{remove\_parallel\_edges}{$\G, E_p$} \;
				\Return $\G$
			}
		\end{algorithm}
	}
	The first phase in Algorithm~\ref{alg:main} is the initialization phase (lines 1 to 9) in which $n$ and $m$ are determined, nodes are created with their required degrees, visited and unvisited sets $I$ and $J$ are created, and $\G$  which includes the maximum degree node is created. 
	The second phase is the tree generation, and third phase is adding the remaining degrees. 
	The fourth and fifth phases are removing the sell loops and parallel edges in order to make $\G$ simple. 

	%\subsubsection*{Adaptive sampling}
	Dynamic adaptive sampling outlined in Algorithm~\ref{alg:sample} is the backbone of the proposed algorithm to satisfy the required degree sequence. It draws a sample $v$ from the input set $I$ by a probability proportional to its nodes' remaining degrees and decrements the degree of $v$ for further samplings. 
	{\SetAlgoNoLine
	\begin{algorithm}[h!]
		\algosize
		\caption{Draw a sample from a set}
		\label{alg:sample}
		\DontPrintSemicolon
		\Function{sample($I$)}{
			draw random $v$ from set $I$ w.r.t. node degrees \;
			decrement $v.d$ \;
			\Return $v$ \;
		}	
	\end{algorithm}
	}
	
	% \subsubsection*{Tree generation}
	Algorithm~\ref{alg:tree} outlines the tree generation algorithm in which at each iteration a random node pair sampled from the visited set $I$ and the unvisited set $J$ are connected. Sampling random nodes from $I$ and $J$ guarantees the connectivity of the output graph $\G$ in which any two nodes are connected by exactly one path, which brings the tree property.
	{\SetAlgoNoLine
	\begin{algorithm}[h]
		\algosize
		\caption{Generate the tree.}
		\label{alg:tree}
		\DontPrintSemicolon
		\Function{generate\_tree($\G, I, J, n$)}{
			\For{$i \gets 1 \ \To \ n-1$}{%
				$u \gets $ \FuncCall{sample}{$I$}  \;
				$v \gets $ \FuncCall{sample}{$J$}  \;
				remove $v$ from $J$ and add it to $I$ \;
				add $v$ to $\G.\V$ \;
				add $(u,v)$ to $\G.\E$ \;
			}
		}
	\end{algorithm}
	}

	% \subsubsection*{Adding remaining edges}
	If there is any remaining edge to be added to $\G$, in other words, if there is any positive degree node pairs in $\G$, it is simply added between these two randomly sampled nodes as summarized in Algorithm~\ref{alg:remain}. Note that self loops and parallel edges are saved in $E_s$ and $E_p$ for later removal.
	{\SetAlgoNoLine
	\begin{algorithm}[h!]
		\algosize
		\caption{Add remaining edges.}
		\label{alg:remain}
		\DontPrintSemicolon
		\Function{add\_remaining\_edges($\G, I, n, m$)}{
			$E_s, E_p \gets [ \ ], [ \ ]$ \;
			\For{$i \gets n \ \To \ m$}{%
				$u \gets $ \FuncCall{sample}{$I$}  \;
				$v \gets $ \FuncCall{sample}{$I$}  \;
				\lIf{$u = v$}{
					append $u$ to $E_s$ 
				}
				\lElseIf{$(u,v) \in \G.\E$}{
					append $(u,v)$ to $E_p$ 
				}
				add $(u,v)$ to $\G.\E$ \;
			}
			\Return{$E_s, E_p$}
		}
	\end{algorithm}
	}
	
\begin{figure}[h!]
		\centering
		\includegraphics[width=0.67\linewidth]{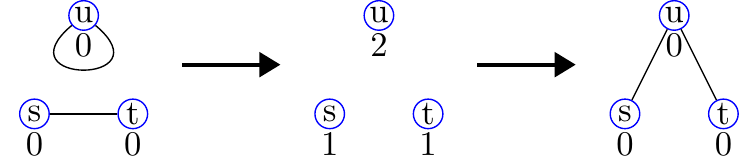}
		\caption{Edge switching operations to remove a self loop}
		\label{fig:self_loop}
	\end{figure}

	To remove the self-loops, we implement the edge switching strategy in Algorithm~\ref{alg:self_loop}.
	Assume that node $u$ has a self loop and other nodes $s$ and $t$ present an edge.
	We first remove the edges $(u,u)$ and $(s,t)$ and then add edges $(u,s)$ and $(u,t)$ to remove the self loop at $u$.
	Note that remaining node degrees do not change after this operation as indicated by the numbers beneath each node in Fig.~\ref{fig:self_loop} which show the number of edges that need  to be added to each node.
	% \subsubsection*{Removing self loops}
	{\SetAlgoNoLine
	\begin{algorithm}[h!]
		\algosize
		\caption{Remove self-loop by edge switching}
		\label{alg:self_loop}
		\DontPrintSemicolon
		\Function{remove\_self\_loops($\G, E_s$)}{
			\While{$|E_s| > 0$}{%
				draw sample $u \in E_s$ \;
				draw sample $(s,t) \in \G$ \;
				\If{$|\{ u, s, t\}| = 3 $ }{
					\If{$(u,s) \not\in G.\E \ \And \ (u,t) \not\in G.\E $ }{
						remove $u$ from $E_s$\;
						remove $(u,u)$ and $(s,t)$ from $\G.\E$\;
						add $(u,s)$ and $(u,t)$ to $\G.\E$\;
					}		
				}
			}
		}
	\end{algorithm}
	}

	Parallel edges are removed similarly as shown in Algorithm~\ref{alg:parallel_edge}.
	Assume $\G$ has parallel edges $(u,v)$ and other nodes $s$ and $t$ present an edge.
	To remove one of the parallel edges between $u$ and $v$, we first remove one of them, i.e.,  $(u,v)$ and $(s,t)$.
	Next, we add edges $(u,s)$ and $(v,t)$ to the graph.
	% It is notable that using this operation, the parallel edges are eliminated without violating the node degree requirement.
	The process of eliminating parallel edges is illustrated in Fig.~\ref{fig:self_loop} where the required node degrees are shown by the numbers beneath each node.
	%Similar to the self-lopp removal, remaining node degrees do no change after this operation as demonstrated by the remaining degrees beneath each node in Fig.~\ref{fig:self_loop}. 
	\begin{figure}[h!]
		\centering
		\includegraphics[width=0.67\linewidth]{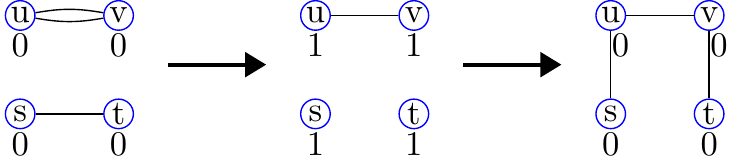}
		\caption{Edge switching operations to remove a parallel edge}
		\label{fig:parallel_edge}
	\end{figure}
	
	% \subsubsection*{Removing parallel edges}
	{\SetAlgoNoLine
	\begin{algorithm}[h!]
		\algosize
		\caption{Remove parallel edges by edge switching}
		\label{alg:parallel_edge}
		\DontPrintSemicolon
		\Function{remove\_parallel\_edges($\G, E_p$)}{
			\While{$|E_p| > 0$}{%
				% remove $(u,v)$ from $\G$\;
				sample $(x,y) \in \G$ \;
				\If{$|\{ u, v, x, y \}| = 4 $ }{
					\If{$(u,x) \not\in \G.\E \ \And \ (v,y) \not\in \G.\E $ }{
						remove $(u,v)$ from $E_p$\;
						remove $(u,v)$ and $(x,y)$ from $\G.\E$\;
						add $(u,x)$ and $(v,y)$ to $\G.\E$\;
					}
					\ElseIf{$(u,y) \not\in \G.\E \ \And \ (v,x) \not\in \G.\E $ }{
						remove $(u,v)$ from $E_p$\;
						remove $(u,v)$ and $(x,y)$ from $\G.\E$\;
						add $(u,y)$ and $(v,x)$ to $\G.\E$\;
					}
				}
			}
		}
	\end{algorithm}
	}

	\subsection{Relabeling the output graph's nodes}
\begin{figure}[h!] % bus wise localization figure
		\centering
		\newcommand{\wdt}{0.25} % figure width
		\subfloat[\label{fig:Gc}]{
			\centering
			\includegraphics[width=\wdt\linewidth]{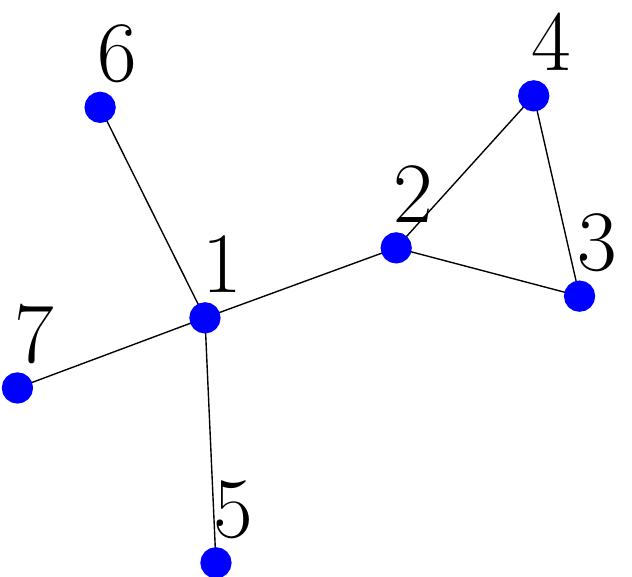}
		}
		\subfloat[\label{fig:Gp}]{
			\centering
			\includegraphics[width=\wdt\linewidth]{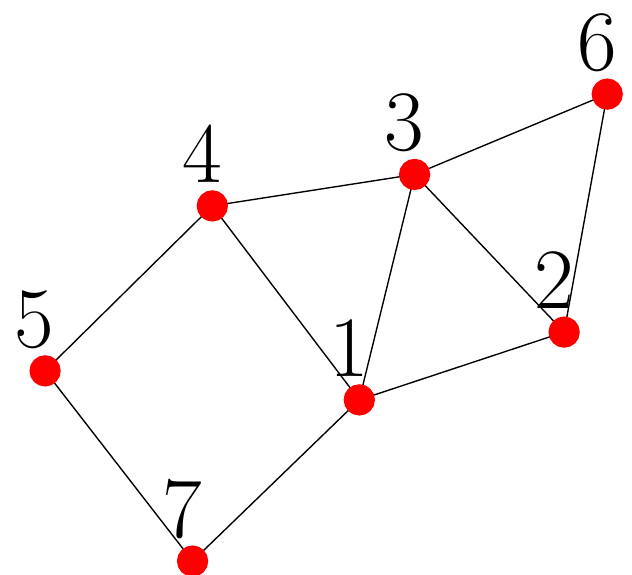}
		}
		\subfloat[\label{fig:Gc2}]{
			\centering
			\includegraphics[width=\wdt\linewidth]{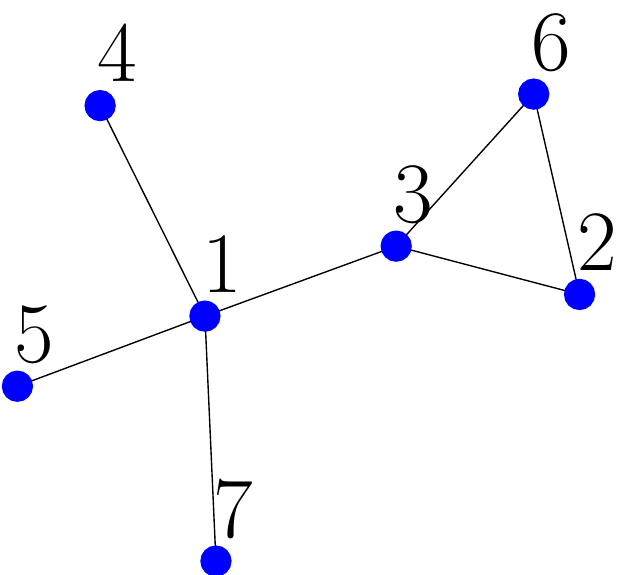}
		}\\
		\caption{Relabeling the generated cyber graph to minimize the cross edge distances. (a) Generated cyber graph. (b) Given power graph. (c) The cyber graph after relabeling. After relabelling, the distance between a power node in (b) and its corresponding cyber node in (c) with the same label is minimized.}
		\label{fig:relabel}
	\end{figure}

	Since the proposed algorithm randomly generates the cyber graph, the nodes' position can be imperfect to match their corresponding nodes in the power graph.
	For instance, a cyber node can be placed far away from the power node that it controls as demonstrated by the cyber node 6 given in Fig.\ref{fig:Gc} and the power node 6 given in Fig.\ref{fig:Gp}. 
	To address this issue, we keep the node positions fixed and relabel the cyber graph labels to minimize the cross links' distances. % between power and cyber nodes. % in order to preserve the cyber graph topology.
	Specifically, we formalize this as an optimization problem \eqref{eq:hungarian} based on the Hungarian algorithm \cite{kuhn1955hungarian} to find the optimal matching between power and cyber nodes: 
	\begin{equation} \label{eq:hungarian}
		\min_{L, R} ,  \textrm{Tr}(LCR) 
	\end{equation}
	where $L$ and $R \in \{0,1\}^{n \times n}$ are the permutation matrices and $C \in R^{n \times n}$ is the cost matrix whose elements are defined by Euclidean distances $C_{u,v} = || u - v ||_2 $ for each $u, v$ pairs from power and cyber graphs, respectively.
	% Figure~\ref{fig:relabel} illustrates the relabeling process for a 7-bus test case.
	% It is clear that, before applying the relabeling process, the node ``4'' in the generated cyber graph matches with power node ``6'' which is not in its vicinity.
	% After applying the relabeling process, the node ``6'' in the generated cyber graph matches with node ``6'' in power graph that minimizes crossing edge distance between two graphs.
	As shown in Figure~\ref{fig:relabel} for a 7-bus test case,  the node ``4'' in the generated cyber graph matches with power node ``6'' which is not in its vicinity before applying the relabeling process.
	After applying the relabeling process, the node ``6'' in the generated cyber graph matches with node ``6'' in power graph that minimizes crossing edge distance between two graphs.

	\section{Results and Discussion} \label{sec:res}
\begin{figure*}[]	% generated graphs
	\centering
	\newcommand{\tdt}{0.018} % text width
	\newcommand{\wdt}{0.28} % figure width
	\setlength{\tabcolsep}{2pt}
	\newcolumntype{?}[1]{!{\vrule width #1}}
	\begin{tabular}{c ?{1pt} c ?{1pt} c ?{1pt} c}
		\textbf{Model} &
		\begin{minipage}[c]{\wdt\textwidth}
			\centering \textbf{30-nodes}
		\end{minipage} &
		\begin{minipage}[c]{\wdt\textwidth}
			\centering \textbf{118-nodes}
		\end{minipage} &
		\begin{minipage}[c]{\wdt\textwidth}
			\centering \textbf{300-nodes}
		\end{minipage}
		\\ \specialrule{1pt}{1pt}{1pt}
		%%%%%%%%%%%%%%%%%%%%%%%%%%%%%%%%%%%%%%%%%%%%%%%%%%%%%%
		\begin{minipage}[c]{\tdt\textwidth}
			\centering \rotatebox{-90}{Configuration model}
		\end{minipage} &
		\begin{minipage}[c]{\wdt\textwidth}
			\includegraphics[width=\linewidth]{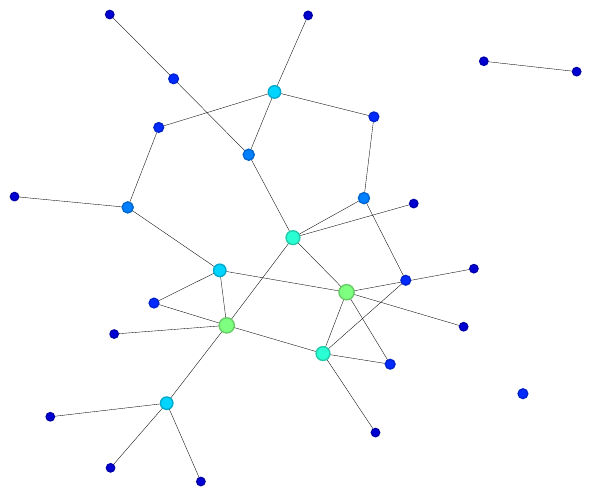}
		\end{minipage} &
		\begin{minipage}[c]{\wdt\textwidth}
			\includegraphics[width=\linewidth]{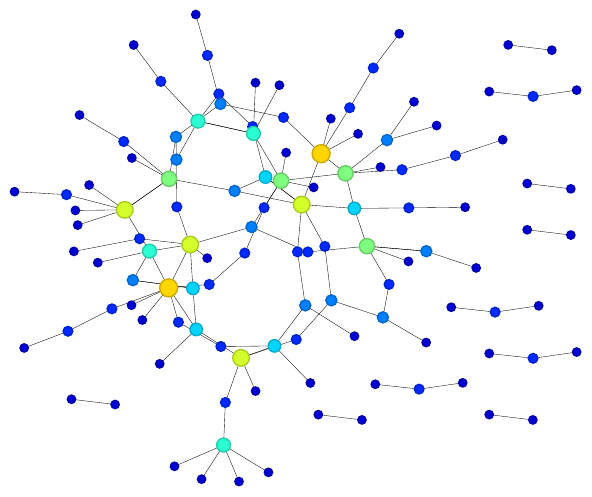}
		\end{minipage} &
		\begin{minipage}[c]{\wdt\textwidth}
			\includegraphics[width=\linewidth]{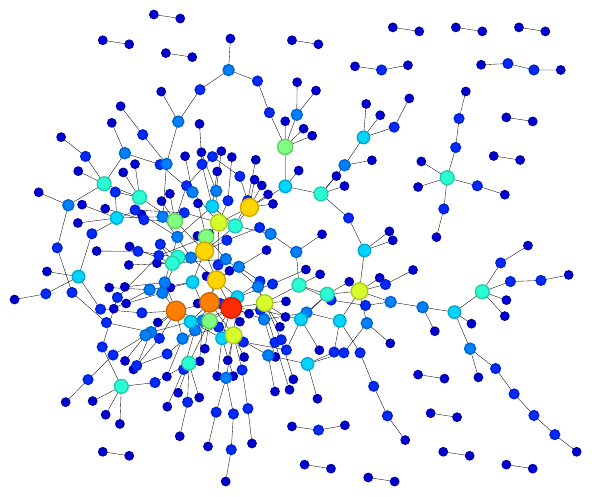}
		\end{minipage} 
		\\ \specialrule{1pt}{1pt}{1pt}
		%%%%%%%%%%%%%%%%%%%%%%%%%%%%%%%%%%%%%%%%%%%%%%%%%%%%%%
		\begin{minipage}[c]{\tdt\textwidth}
			\centering \rotatebox{-90}{Havel-Hakimi}
		\end{minipage} &
		\begin{minipage}[c]{\wdt\textwidth}
			\includegraphics[width=\linewidth]{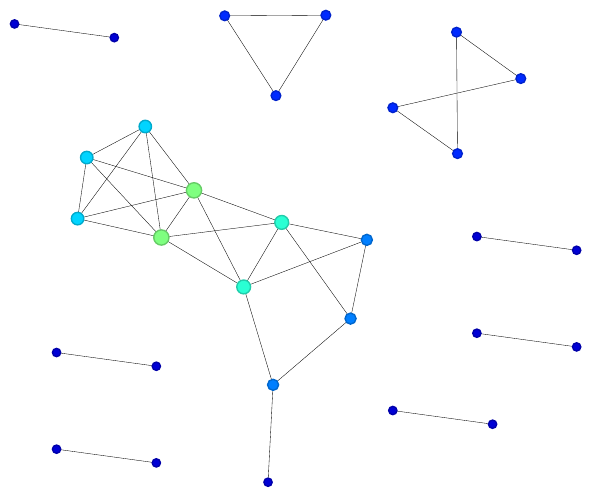}
		\end{minipage} &
		\begin{minipage}[c]{\wdt\textwidth}
			\includegraphics[width=\linewidth]{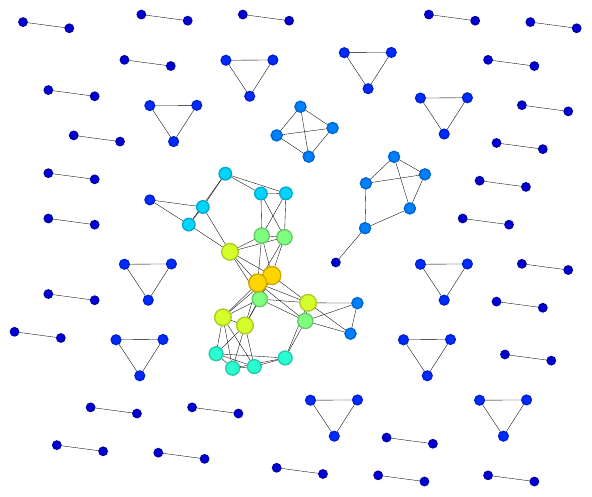}
		\end{minipage} &
		\begin{minipage}[c]{\wdt\textwidth}
			\includegraphics[width=\linewidth]{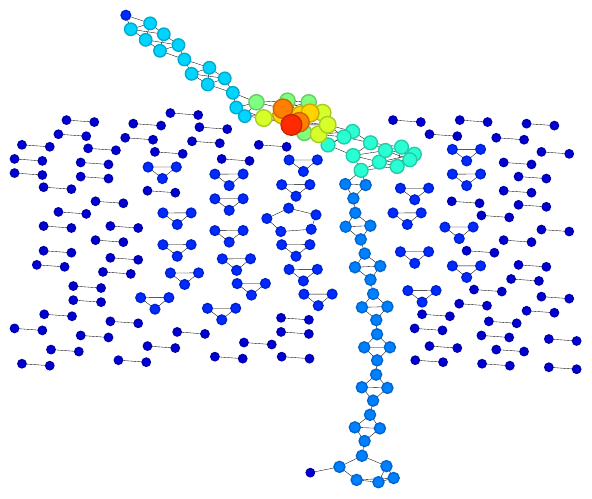}
		\end{minipage} 
		\\ \specialrule{1pt}{1pt}{1pt}
		%%%%%%%%%%%%%%%%%%%%%%%%%%%%%%%%%%%%%%%%%%%%%%%%%%%%%%
		\begin{minipage}[c]{\tdt\textwidth}
			\centering \rotatebox{-90}{Chung-Lu}
		\end{minipage} &
		\begin{minipage}[c]{\wdt\textwidth}
			\includegraphics[width=\linewidth]{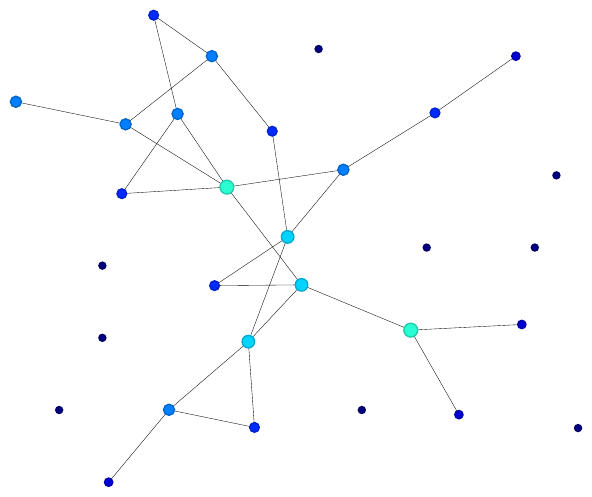}
		\end{minipage} &
		\begin{minipage}[c]{\wdt\textwidth}
			\includegraphics[width=\linewidth]{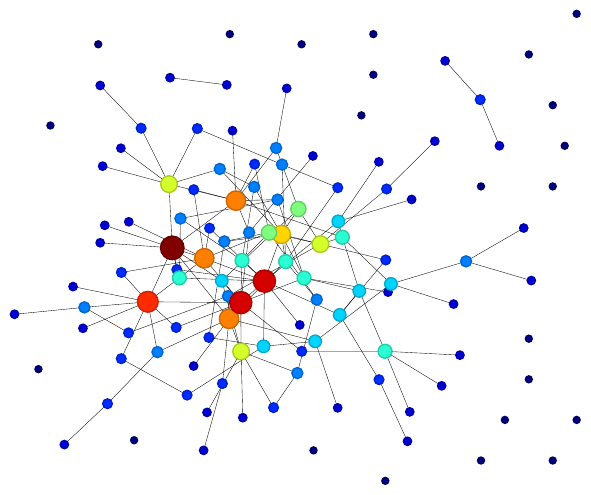}
		\end{minipage} &
		\begin{minipage}[c]{\wdt\textwidth}
			\includegraphics[width=\linewidth]{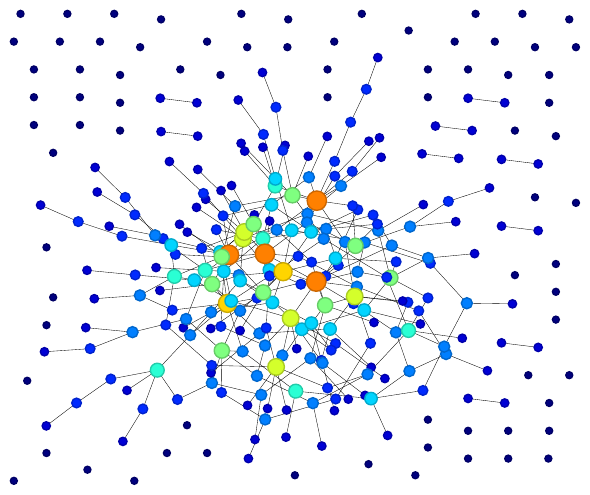}
		\end{minipage} 
		\\ \specialrule{1pt}{1pt}{1pt}
		%%%%%%%%%%%%%%%%%%%%%%%%%%%%%%%%%%%%%%%%%%%%%%%%%%%%%%
		\begin{minipage}[c]{\tdt\textwidth}
			\centering \rotatebox{-90}{Horv\'{a}t-Modes}
		\end{minipage} &
		\begin{minipage}[c]{\wdt\textwidth}
			\includegraphics[width=\linewidth]{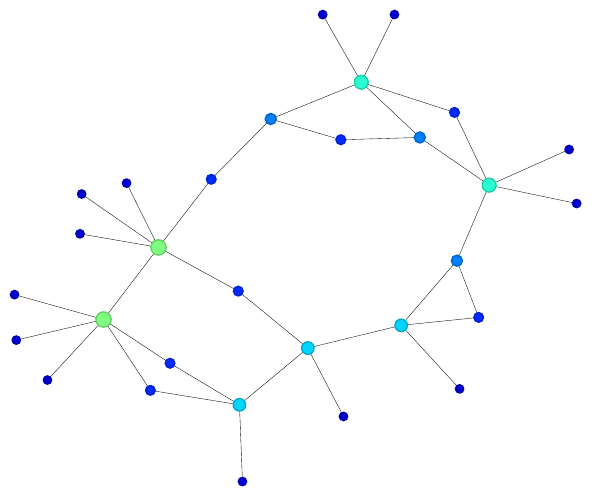}
		\end{minipage} &
		\begin{minipage}[c]{\wdt\textwidth}
			\includegraphics[width=\linewidth]{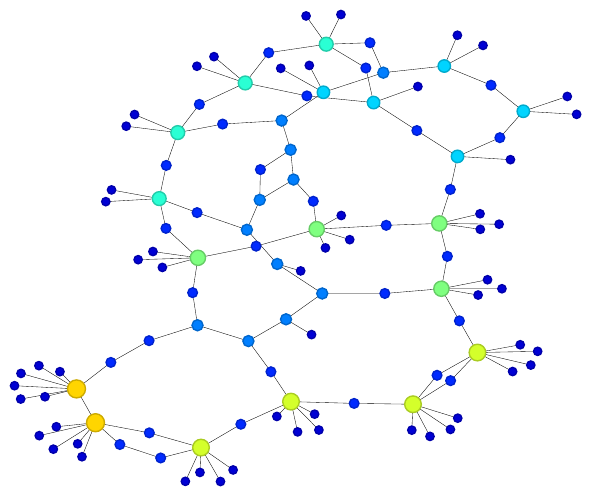}
		\end{minipage} &
		\begin{minipage}[c]{\wdt\textwidth}
			\includegraphics[width=\linewidth]{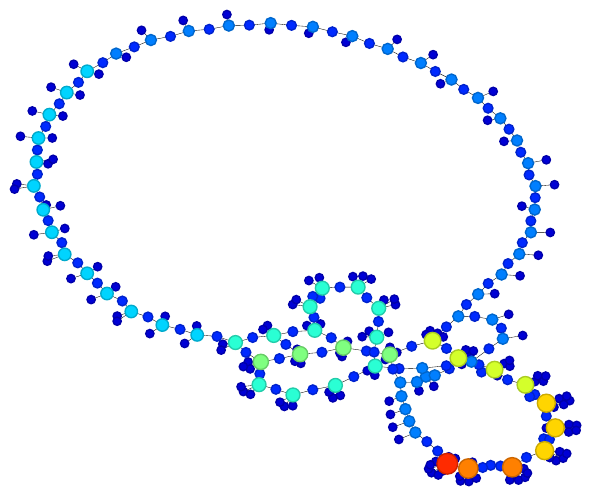}
		\end{minipage} 
		\\ \specialrule{1pt}{1pt}{1pt}
		%%%%%%%%%%%%%%%%%%%%%%%%%%%%%%%%%%%%%%%%%%%%%%%%%%%%%%
		\begin{minipage}[c]{\tdt\textwidth}
			\centering \rotatebox{-90}{Proposed work}
		\end{minipage} &
		\begin{minipage}[c]{\wdt\textwidth}
			\includegraphics[width=\linewidth]{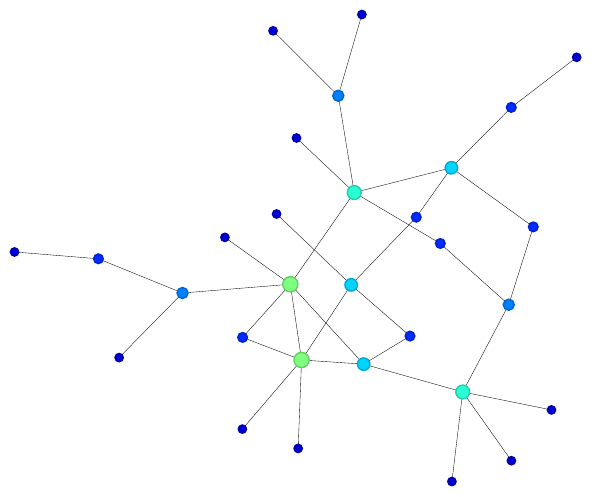}
		\end{minipage} &
		\begin{minipage}[c]{\wdt\textwidth}
			\includegraphics[width=\linewidth]{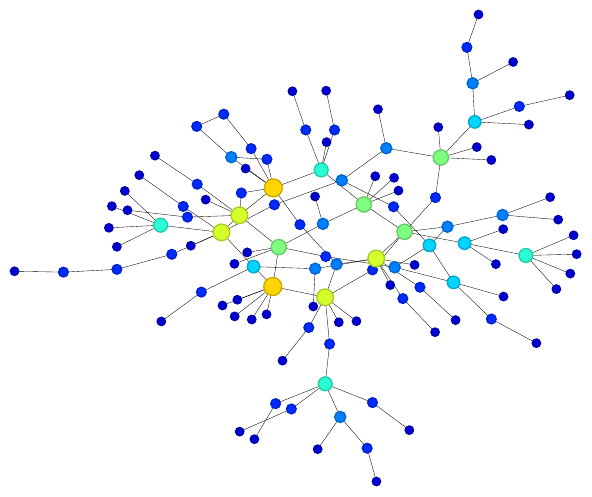}
		\end{minipage} &
		\begin{minipage}[c]{\wdt\textwidth}
			\includegraphics[width=\linewidth]{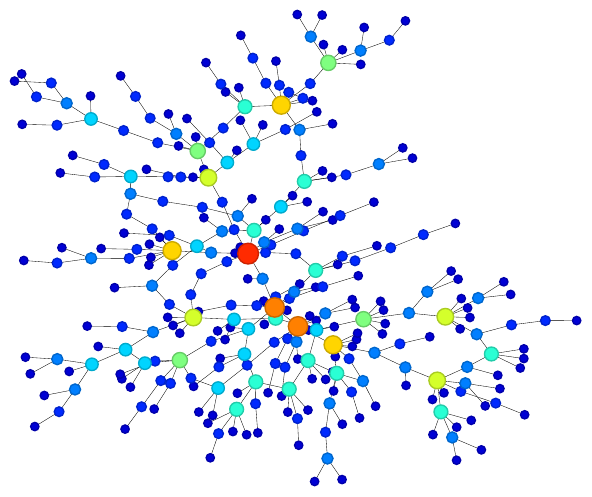}
		\end{minipage} 
		\\ %\hlines
	\end{tabular}
	\includegraphics[width=0.97\linewidth]{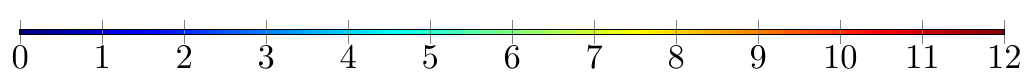}
	\caption{Generated graphs for each model (rows) and system (columns). Node's colors and sizes are given according to their degrees for better visualization}
	\label{fig:graphs}
	\end{figure*}

   	For a fair comparison, we implemented the existing approaches in the literature such as configuration model (CM), Havel-Hakimi (HH), Chung-Lu (CL), and Horv\'{a}t-Modes (HM) algorithms.
  	Then, we generate random graphs for 30-, 118-, and 300-buses IEEE test systems using the same degree sequence.
  	Generated graphs are visualized in Fig.~\ref{fig:graphs} where nodes' color and sizes indicate their degrees. In addition, the global characteristics of these graphs are tabulated in Table~\ref{tab:chars}. 
   	
	% global graph charachteristics
	\begin{table}[h!] 
	\centering
	\setlength{\tabcolsep}{4pt}
	\renewcommand{\arraystretch}{1.4}
	\newcolumntype{?}[1]{!{\vrule width #1}}
	\caption{Global graph characteristics for each model and system (CM:Config. model, HH: Havel-Hakimi, CL:Chung-Lu HM:Horv\'{a}t-Modes, PW:Proposed work}
	\begin{tabular}{c?{1pt} c | c | c | c | c | c | c | c | c}
		\textbf{n} & model & $\G.g$ & $\G.c$ & $\rho$ & $\diameter $ & $\widetilde{sp}$ & $cc$   &  $a$  & $\lambda$\\
		\specialrule{1pt}{1pt}{1pt}
		\multirow{4}{*}{\textbf{30}}  
		& CM   & \xmark  & \xmark & 1.167 & \NA & \NA  &  \NA  & -0.128 & $9.62e^{-2}$		\\ \cline{2-10}
		& HH   & \cmark  & \xmark & 1.167 & \NA & \NA  & 0.303 &  0.776 & $2.19e^{-1}$		\\ \cline{2-10}
		& CL   & \cmark  & \xmark & 0.933 & \NA &  \NA & 0.098 & -0.132 & $9.80e^{-2}$		\\ \cline{2-10}
		& HM   & \cmark  & \cmark & 1.167 &   8 & 3.84 & 0.050 & -0.676 & $4.74e^{-2}$		\\ \cline{2-10}
		& \textbf{PW} & \cmark  & \cmark & 1.167 &   7 & 3.44 & 0.048 & -0.220 & $8.62e^{-2}$ 		\\ 
		\specialrule{1pt}{1pt}{1pt}
		
		\multirow{4}{*}{\textbf{118}}  
		& CM   & \xmark  & \xmark & 1.102 & \NA & \NA  &  \NA  & -0.010 & $3.13e^{-2}$		\\ \cline{2-10}
		& HH   & \cmark  & \xmark & 1.102 & \NA & \NA  & 0.417 &  0.880 & $1.22e^{-1}$		\\ \cline{2-10}
		& CL   & \cmark  & \xmark & 1.220 & \NA & \NA  & 0.058 & -0.002 & $8.86e^{-2}$		\\ \cline{2-10}
		& HM   & \cmark  & \cmark & 1.102 &  17 & 7.88 & 0.000 & -0.737 & $7.68e^{-3}$		\\ \cline{2-10}
		& \textbf{PW} & \cmark  & \cmark & 1.102 &  13 & 5.64 & 0.026 & -0.220 & $1.68e^{-2}$ 		\\ 
		\specialrule{1pt}{1pt}{1pt}
		
		\multirow{4}{*}{\textbf{300}}  
		& CM   & \xmark  & \xmark & 1.040 & \NA & \NA  &  \NA  & -0.059 & $7.77e^{-3}$		\\ \cline{2-10}
		& HH   & \cmark  & \xmark & 1.040 & \NA & \NA  & 0.365 &  0.927 & $1.79e^{-3}$		\\ \cline{2-10}
		& CL   & \cmark  & \xmark & 0.943 & \NA & \NA  & 0.008 & -0.029 & $4.50e^{-2}$		\\ \cline{2-10}
		& HM   & \cmark  & \cmark & 1.040 &  54 &21.22 & 0.000 & -0.606 & $7.62e^{-4}$		\\ \cline{2-10}
		& \textbf{PW} & \cmark  & \cmark & 1.040 &  22 & 9.22 & 0.007 & -0.226 & $2.86e^{-3}$ 		\\ 
		% \specialrule{1pt}{1pt}{1pt}
	\end{tabular}
	% \vspace{1mm}
	% \\CM: Config. model, HH: Havel-Hakimi, CL: Chung-Lu \\ HM: Horv\'{a}t-Modes, PW: Proposed work
	\label{tab:chars}
	\end{table}   	
   	
   	As can be seen from the Fig.~\ref{fig:graphs} and Table~\ref{tab:chars}, although these graphs are generated by the same degree distribution, their configurations are totally different.
   	The graph generated by ``Configuration Model'' is neither graphical, nor connected. 
   	Havel-Hakimi model, in contrast, generates graphical outputs.
   	However, this model does not guarantee connectivity of the graph and also it has a high clustering coefficient $(cc>0.3)$ and assortativity $(a>0.7)$.
    These features makes Havel-Hakimi model ineligible for generating realistic cyber graphs. 
	Similar to the Havel-Hakimi model, Chung-Lu model generates graphical but not connected graphs.
	However, since it utilizes the expected degree distribution instead of the exact one, its generated graphs may show different characteristics from the graph that generated by the exact degree distribution.
	
	For instance, the graphs generated by the Chung-Lu model for IEEE 30- and 300-bus test systems present less edges $(\rho<0.95)$ than the required edges.
	Conversely, the graph generated by the Chung-Lu model for the 118-bus test system has more edges $(\rho>1.2)$ compared to the required one.
	Horv\'{a}t-Modes model, contrary to the previously mentioned models, generates both graphical and connected outputs.
	Yet, its highly low assortativity $(a<-0.6)$ is not compatible with the real-world cyber graph.
	In addition, the graph generated for 300-bus test system presents  high diameter $(\diameter=54)$ and high average shortest path ($\widetilde{sp}>21$) which indicate its ``bias'' induced to preserve the connectivity. 
	In contrast, our proposed method generates tree-like connected structures that have the exact degree distribution and similar graph attributes to the reference system. % given in \ref{tab:ref_char}.
	Moreover, it generates a graph which has a diameter and an average shortest paths proportional to its node size, an appropriate clustering coefficient, a spectral gap decreasing with its node size, and almost the same assortativity $(a=-0.22)$ with the reference system.
	
	Another finding  of our experiments is that although the graphs generated by Havel-Hakimi, Horv\'{a}t-Modes, and the proposed work exhibit the  same distribution, they produce totally different topologies, especially for the 300-bus test system in Fig.~\ref{fig:graphs}.
	For instance, Havel-Hakimi produces many 2-vertex or 3-vertex disconnected cliques.
	Horv\'{a}t-Modes, on the contrary, tends to create graphs with higher assortativity to preserve the connectivity.
	In contrast, the proposed model generates tree like structures to better model a real-world cyber graph.
	Aside from the global graph characteristics given in Table~\ref{tab:chars}, this situation  corroborates the supremacy of the proposed algorithm over the existing ones.
	
    \section{Conclusion} \label{sec:con}	
	In this paper, we propose a simple and elegant framework by modifying the Chung-Lu algorithm to generate connected, simple, and realistic cyber graphs for power systems.
	For generating graphs which have the exact required node degree distribution, we propose an adaptive remaining degree approach instead of the fixed expected degree in the Chung-Lu method.
	In addition, we utilize the Hungarian algorithm to minimize the length of cross-links between the power graph and the generated cyber graph.
	We implement other graph generation methods to compare the suitability of the proposed algorithm with those in literature including configuration model, Havel-Hakimi, Chung-Lu, and Horv\'{a}t-Modes algorithms and generate cyber graphs using the same degree sequence for each model.
	Generated cyber graphs for IEEE 30-, 118-, and 300-bus test systems demonstrate that the proposed model yields better results compared to the existing approaches in terms of global characteristics, connectivity, and graphicality to mimic a real-world communication system.
	The proposed framework can be utilized by power system community to generate realistic cyber graphs for cyber-physical power system studies.
	
	\bibliographystyle{IEEEtran}
	\bibliography{tpec2022}

\end{document}